\documentclass[aps,prb,preprint]{revtex4}
\usepackage{graphicx,psfrag}
\usepackage{graphicx} 
\usepackage{bm}
\usepackage{amsmath,amsthm,amssymb}
\newlength{\minitwocolumn}
\begin{document}

\title{
The anomalous Hall conductivity due to the vector spin chirality 
}

\author{Katsuhisa Taguchi}  

\author{Gen Tatara}

\affiliation{%
Department of Physics, Tokyo Metropolitan University,
Hachioji, Tokyo 192-0397, Japan
}

\date{\today}
%
\begin {abstract} 
We study theoretically the anomalous Hall effect due to the vector spin chirality carried by the local spins in the $s$-$d$ model.
We will show that the vector spin chirality indeed induces local Hall effect in the presence of the electron spin polarization, 
while the global Hall effect vanishes if electron transport is homogeneous.
This anomalous Hall effect can be interpreted in terms of the rotational component of the spin current associated with the vector chirality.

\end{abstract}

\maketitle

\section{Introduction} 

The anomalous Hall effect is the Hall effect due to the magnetization in ferromagnets. 
This anomalous contribution is usually attributed to the spin-orbit interaction, and is usually proportional to the magnetization\cite{rf:Kapulus_54,rf:Smit_58,rf:Berger_70,rf:Kondo_62}.
Recently, it was found that some manganese ferromagnetic pyrochlores can exhibit abnormal anomalous Hall effects\cite{rf:Ye_99, rf:Y. Taguchi_01} different from the conventional ones.  
The behavior was due to the spin Berry phase associated with the non-trivial spin configuration (spin chirality) driven thermally or by the geometrical frustration \cite{rf:Ye_99, rf:Ohgushi_00, rf:Y. Taguchi_01}.  Such anomalous Hall effect was first pointed out in the strong Hund-coupling limit\cite{rf:Ye_99}, considering a half-metallic nature of the experimental systems.  The weak coupling limit was studied by applying the linear response theory and the perturbation expansion with respect to the $s$-$d$ interaction\cite{rf:Tatara_02}. In the perturbation regime, anomalous Hall effect was found to be induced by the scalar spin chirality,  $\bm{S}_i\cdot (\bm{S}_j\times \bm{S}_k)$, which is made from three localized spin $\bm{S}_i$, $\bm{S}_j$, and $\bm{S}_k$ $(i, j$, and $k$ represent the position). 
The spin chirality is odd under the time reversal and even under the space inversion. 
The scalar chirality becomes a nonzero and gives rise a large anomalous Hall effect when the spin configuration is noncoplanar.
The chirality reduces to the spin Berry phase in the slowly varying spin structures\cite{rf:M.Onoda_04}.

The spin Berry phase and scalar chirality have been shown to induce spontaneous charge current\cite{rf:Loss_91,rf:Tatara_03}.  This circulating current was pointed out to be the origin of the Hall effect by the spin scalar chirality\cite{rf:Tatara_03}. In the context of spintronics, another spin chirality that is associated with the spin current is of particular interest. 
This is the vector spin chirality, $\bm{S}_i \times \bm{S}_j$, carried by non-collinear local spins. The vector chirality is even under the time reversal and odd under the space inversion. 
The vector spin chirality has shown to drive electric polarization in Mott insulators through the spin-orbit interaction such as multi-ferroic manganese oxides\cite{rf:kimura_03,rf:Katsura_05,rf:Mostovoy_06,rf:Sergienko_06}. 

The aim of the paper is to investigate the role of the vector spin chirality in the anomalous Hall effect in metals. 
The Hall conductivity is calculated based on the $s$-$d$ model taking account of vector spin chirality perturbatively. 
We consider the case where the conductive electron has the uniform spin polarization and in the presence of the scattering by non-magnetic impurities. The spin-orbit interaction is not taken account of. The localized spin are treated as classical and static. We demonstrate that the vector chirality indeed drives the local Hall effect if the conductive electron is uniformly spin polarized.

\section{Hall conductivity} 
The Hamiltonian we consider is written as  $\mathcal{H}_{0}+\mathcal{H}_{\mathrm{sd}}+\mathcal{H}_{\mathrm{em}}$, where
\begin{align}\label{eq:H-free}  
&\mathcal{H}_{0}=\sum_{k, \sigma=\pm} \epsilon_{k,\sigma}c_{k, \sigma}^{\dagger}(t)c_{k, \sigma}(t)
\\ \label{eq:H-sd}    
&\mathcal{H}_{\mathrm{sd}}=
-{J_\mathrm{sd}}\sum_{kk'}{\bm{S}_{\bm{k'}-\bm{k}}}\cdot (c_{k'}^{\dagger}(t){\bm\sigma} c_{k}(t)),
\end{align}
are the free electron part and the $s$-$d$ interaction respectively. 
Here $\sigma=\pm$ denotes the electron spin and $c_{k, \sigma}(t)$ and $c_{k, \sigma}^{\dagger}(t)$ are the electron annihilation and creation operators $(c_{k}(t)\equiv (c_{k,+}(t), c_{k,-}(t)))$. 
The conductive electron energy is $\epsilon_{k, \sigma}(\equiv \frac{\hbar^2 k^2}{2m}-\epsilon_{\mathrm{F}}-\sigma M)$,  where $m$ is the electron mass, $M$ is the uniform spin polarization of conduction electron(due to the magnetization or the external field), and $\epsilon_{\mathrm{F}}$ is the Fermi energy. 
Uniform polarization is chosen as along the $z$ axis. Localized spin at the position $\bm{x}$ is written as $\bm{S}(\bm{x})=\sum_k e^{i\bm{k}\cdot\bm{x}}\bm{S}_k$, and ${\sigma ^{\alpha}}(\alpha =x,y,z)$ are the 2 $\times$ 2 Pauli matrices.   

The effect of the impurity scattering is taken account of by the lifetime, $\tau$, of the electrons. 
The term $\mathcal{H}_{\mathrm{em}}$ represents the interaction with the applied electric field, $\mathcal{H}_{\mathrm{em}}=-\int d\bm{x} \bm{j}(\bm{x})\cdot\bm{a}(\bm{x},t)$, where $\bm{a}(\bm{x}, t)$ is the U(1) vector potential, related to the applied electric filed. 
Assuming the case of spatially uniform electric field, the $\mathcal{H}_{\mathrm{em}}$ is written as 
\begin{align}
\mathcal{H}_{\mathrm{em}}=-\sum_{k, \nu ,\sigma=\pm}\frac{i e^{i\Omega t}}{\Omega}\frac{e\hbar}{m}k^{\nu}E^{\nu} c_{k, \sigma}^{\dagger}(t)c_{k, \sigma}(t),
\end{align}
where $\Omega$ is the frequency of the external electric field. Considering the static Hall conductivity, $\Omega$ is chosen as zero at the end of the calculation. The charge current density is defined as
\begin{align} \label{eq:charge current}
\bm{j}(\bm{x})=&\frac{\mathrm{e} \hbar^2}{2i mV}\sum_{k, k'}\int \frac{\mathrm{d}\omega}{2\pi}
\left(\bm{k}+\bm{k'}	\right)e^{i(\bm{k'-k})\cdot\bm{x}}\mathrm{Tr}\left[G^<_{k, k',\omega}\right], 
\end{align}
where $\mathrm{Tr}$ represents trace over the spin indices and $G^{<}_{k,k',\omega}$ is the Fourier transform  of the lesser Green's function defined as $G^{<}_{k,k',t}\equiv (i\hbar)^{-1} \langle  c^{\dagger}_{k}(t) c_{k'}(t) \rangle$ ($G^{<}$ is  a $2\times2$  matrix in the spin space), with $\omega$ being the frequency of the electron. We include the exchange interaction perturbatively to the second order, considering the weak exchange interaction, $J_{\mathrm{sd}}  \ll \hbar/\tau \ $\cite{rf:Tatara_02}.
\begin{figure}
\centering
\includegraphics[scale=0.8]{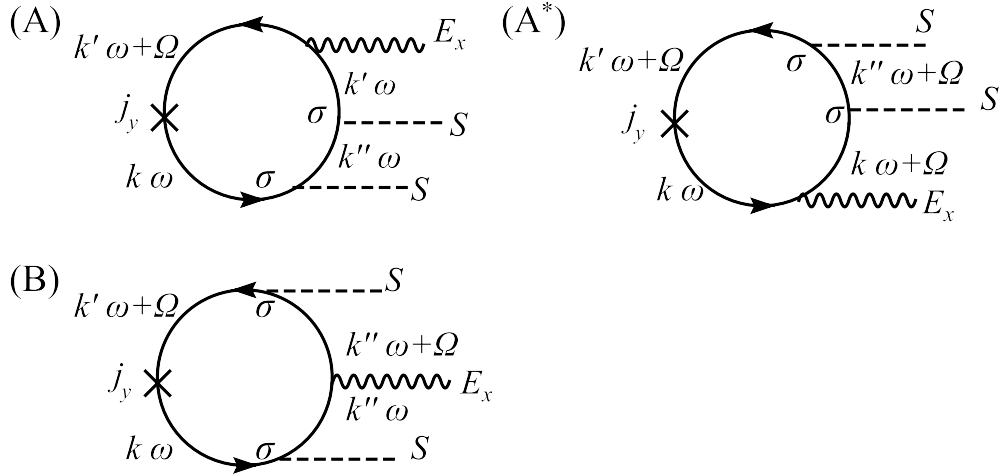}
\caption{Vector chirality contribution to the Hall conductivity in the liner order in the applied field, $E_{\mathrm{x}}$, represented by the wavy lines.  Dotted lines represent the interaction with local spins, $\bm{S}$. 
}\label{fig:1}
\end{figure}

We calculate the local (i.e., $\bm{x}$-dependent) current density along the $y$-direction. The electric field is applied in the $x$-direction. The contribution to the current is diagrammatically shown in Fig.1.  The first contribution (A) in Fig.1 is written as  
\begin{align} \notag
j^{(\mathrm{A})}_{y}(\bm{x}) =&-\lim_{\Omega \to 0}\frac{\hbar}{2V\Omega}J_{\mathrm{sd}}^2 E_{x}\left(\frac{e \hbar}{m}\right)^2
\sum_{kk'k''}\sum_{\alpha, \beta} \int \frac{\mathrm{d}\omega}{2\pi}(k+k')_{y} k'_{x}E_{x}  S^{\alpha}_{k-k''}S^{\beta}_{k''-k'}e^{i(\bm{k'}-\bm{k})\cdot \bm{x}}
\\ \label{eq:calculatio-a}
&\times  \mathrm{Tr}	[G_{k,\omega}\sigma_{\alpha}G_{k'',\omega}\sigma_{\beta}G_{k',\omega}G_{k',\omega+\Omega}]^{<}.
\end{align}
 
The lesser component can be expanded by using retarded and advanced Green's function as\cite{rf:book1} 
\begin{align}\notag
&[G_{k,\omega}\sigma_{\alpha}G_{k'',\omega}\sigma_{\beta}G_{k',\omega}G_{k',\omega+\Omega}]^{<}
\\ \notag
&=(f(\omega+\Omega)-f(\omega)){G^R_{k,\omega}\sigma_{\alpha}}G^R_{k'',\omega}\sigma_{\beta}G^R_{k',\omega}G^A_{k',\omega+\Omega}
\\ \label{eq:keldysh}
&+f(\omega+\Omega){G^R_{k,\omega}\sigma_{\alpha}G^R_{k'',\omega}\sigma_{\beta}G^R_{k',\omega}G^R_{k',\omega+\Omega}}-f(\omega){G^A_{k,\omega}\sigma_{\alpha}G^A_{k'',\omega}\sigma_{\beta}G^A_{k',\omega}G^A_{k',\omega+\Omega}},
\end{align}
where $f(w)$ is the Fermi distribution function is given as $f(w)=-\theta(w)$ at zero temperature ($\theta(w)$ is the step function). 
We neglect the terms containing only $G^R$'s or $G^A$'s,  which are higher order of $\frac{\hbar}{\epsilon_{\mathrm{F}}\tau}\ll 1$ compared with the contribution of the $G^R$ and $G^A$ mixed.
Then Eq. (\ref{eq:keldysh}) is written as,
\begin{align}\label{eq:w}
&[G_{k,\omega}\sigma_{\alpha}G_{k'',\omega}\sigma_{\beta}G_{k',\omega}G_{k',\omega+\Omega}]^{<}
\simeq -\Omega\delta(\omega){G^R_{k}\sigma_{\alpha}G^R_{k''}\sigma_{\beta}G^R_{k'}G^A_{k'}}.
\end{align}
Here $\mathrm{G}^R_{k}$ and  $\mathrm{G}^A_{k} \left( =\left( \mathrm{G}^R_{k} \right)^* \right)$ are retarded and advanced Green's functions with zero frequency. The Green's functions include the effect of the uniform magnetization, and so are $2 \times 2$ diagonal matrices, e.g., 
\begin{align}
        &G^R_{k }=\left(\begin{array}{cc}
		G^R_{k + } & 0 \\
		0 & G^R_{k - }
\end{array}\right),
\end{align} 
where the components are given as $G^{R}_{k\sigma}=(-\epsilon_{k{\sigma}}+\frac{i\hbar}{2\tau})^{-1}$. From Eqs. (\ref{eq:calculatio-a}) and (\ref{eq:w}), the contribution from the first and second diagrams of Fig.1 $(\mathrm{A}$ and $\mathrm{A^*})$ to the local Hall conductivity is obtained as,
\begin{align} \notag
\sigma_{xy}^{(\mathrm{A}+\mathrm{A}^*)}(\bm{x})
=&\frac{\hbar J_{\mathrm{sd}}^2}{4\pi V}\left(\frac{e \hbar}{m}\right)^2 \sum_{kk'k''}\sum_{\alpha, \beta}k_{y} k'_{x} 	S^{\alpha}_{k-k''}S^{\beta}_{k''-k'}e^{i(\bm{k'}-\bm{k})\cdot \bm{x}} 
\mathrm{Tr}[{G^R_{k}\sigma_{\alpha}}G^R_{k''}\sigma_{\beta}G^R_{k'}G^A_{k'}+\mathrm{c.c}],
\\ \label{eq:calculatio-a2}
=&\iint d\bm{x}_{1}d\bm{x}_{2} A^{(\mathrm{A})}(\bm{x}, \bm{x}_{1}, \bm{x}_{2})\ (\bm{S}_{1} \times \bm{S}_{2})^z,
\end{align}
where
\begin{align}\notag
A^{(\mathrm{A})}(\bm{x}, \bm{x}_{1}, \bm{x}_{2})\equiv& \frac{\hbar J_{\mathrm{sd}}^2}{2\pi V^3} \left(\frac{e \hbar}{m}\right)^2 \sum_{\sigma}\sigma
\\ \label{eq:coefficient-a}
\times \mathrm{Im}&\left[\sum_{k}\partial_{x_{10}} e^{-i\bm{k}\cdot \bm{x}_{10}} G^R_{k,\sigma}\sum_{k'}\partial_{y_{20}}e^{i\bm{k'}\cdot \bm{x}_{20}}|G^R_{k',\sigma}|^2\sum_{k''}e^{i\bm{k''}\cdot \bm{x}_{12}}G^R_{k'',-\sigma}\right].
\end{align}
Here ${\bm{S}}_i\equiv{\bm{S}}({\bm{x}}_i), (i=1, 2)$ are the localized spins in the real space, and ${\bm{x}_{i0}}\equiv{\bm{x}_{i}}-{\bm{x}}$, ${\bm{x}_{12}}\equiv \bm{x}_{1}- \bm{x}_{2}$. Similarly, we obtain the contribution $(\mathrm{B})$ in Fig. 1 as 
\begin{align}\label{eq:calculatio-b2}
\sigma_{xy}^{(\mathrm{B})}(\bm{x})&=\iint d\bm{x}_{1}d\bm{x}_{2}A^{(\mathrm{B})}(\bm{x}, \bm{x}_{1}, \bm{x}_{2})\ (\bm{S}_{1} \times \bm{S}_{2})^z,
\end{align}
where
\begin{align} \notag
&A^{(\mathrm{B})}(\bm{x}, \bm{x}_{1}, \bm{x}_{2}))=
-\frac{ \hbar J_{\mathrm{sd}}^2}{2\pi V^3} \left(\frac{e \hbar}{m}\right)^2\sum_{\sigma}\sigma
\\ \label{eq:result-b}
&\times \mathrm{Im} \left[ \sum_{k} \partial_{y_{10}}e^{-i\bm{k}\cdot \bm{x}_{10}} G^R_{k,\sigma}\sum_{k'}e^{i\bm{k'}\cdot \bm{x}_{20}}G^A_{k',\sigma}\sum_{k''}\partial_{x_{12}}e^{i\bm{k''}\cdot \bm{x}_{12}}|G^R_{k'',-\sigma}|^2 \right].
\end{align}
The total local Hall conductivity, $\sigma_{xy}(\bm{x})=\sigma_{xy}^{(\mathrm{A}+\mathrm{A}^*)}(\bm{x})+\sigma_{xy}^{(\mathrm{B})}(\bm{x})$, is therefore obtained as 
\begin{align}\label{eq:Hall conducutivity-1}
\sigma_{xy}(\bm{x})&=\iint d\bm{x}_{1}d\bm{x}_{2}
A_{12}(\bm{x})\ (\bm{S}_{1} \times \bm{S}_{2})^z,
\end{align}
where
\begin{align} \notag
&A_{12}(\bm{x})=\frac{- J_{\mathrm{sd}}^2\tau}{\pi  V^3}\left(\frac{e \hbar}{m}\right)^2(\bm{x}_{10}\times \bm{x}_{20} )_{\mathrm{z}}
\\ \notag
&\times\sum_{\sigma}\sigma \ \mathrm{Im}\left[i
\frac{I'_{\sigma}(r_{10}){
\mathrm{Im}\left[ I'_{\sigma}(r_{20})\right]I_{-\sigma}(r_{12})
}}{ r_{10}r_{20}}
+i\frac{I'_{\sigma}(r_{10})\mathrm{Im}\left[ I'_{-\sigma}(r_{12})\right]I^{*}_{\sigma}(r_{20})}{ r_{10}r_{12}}\right].
\\ \label{eq:A+B}
\end{align}
The correlation function is given as $I_{\sigma}(r_{ij})=\sum_{k}e^{i\bm{k}\cdot\bm{x}_{ij}}
G^R_{k, \sigma}\simeq -\frac{N_{e_{\sigma}}\pi}{k_{\mathrm{F}_{\sigma}}r_{ij}}e^{-\frac{r_{ij}}{2\ell_{\sigma}}}
e^{ik_{\mathrm{F}_{\sigma}}r_{ij}}$, ($k_{\mathrm{F}_{\sigma}}\equiv(1-\sigma\frac{M}{2\epsilon_\mathrm{F}})k_\mathrm{F}$) 
and  $I' (r)$ denotes its derivative. Here $N_{e_{\sigma}}=N_{e} \sqrt{1-\sigma\frac{M}{\epsilon_{\mathrm{F}}}}$, 
($N_{e}=\frac{mVk_{\mathrm{F}}}{2\pi \hbar^2 }$) 
is the spin-dependent density of states.
The function $I(r_{ij})$ depends only on the distance, $r_{ij}\equiv|\bm{x_{ij}}|$.  It oscillates with wave  length of $k_{\mathrm{F}\sigma}$ and decreases within the distance of the mean free path, $\ell_{\sigma} \equiv \frac{\hbar k_{\mathrm{F}_{\sigma}}\tau}{m}$. 
When $\frac{M}{\epsilon_{\mathrm{F}}}$ is small and neglecting higher orders of $({k_{\mathrm{F}}\ell})^{-1}$, the coefficient $A_{12}(\bm{x})$ is estimated as 
\begin{align}\label{eq:BC1}
A_{12}(\bm{x})\simeq
&\left(\frac{(\bm{x}_{10}\times \bm{x}_{20})^z}{r_{10}r_{20}}+ \frac{(\bm{x}_{12}\times \bm{x}_{10})^z}{r_{10}r_{12}} \right)\left(\frac{e \hbar}{m}\right)^2 \frac{J_{\mathrm{sd}}^2 M {N}^3_{e}\tau }{4\epsilon_{\mathrm{F}} V^3} \frac{1}{r_{10}^2}  e^{-\frac{r_{10}+r_{20}+r_{12}}{2\ell}}.
\end{align}

The local Hall effect vanishes in the absent of spin polarization, as seen from the fact that $A_{12}(\bm{x})=0$ if $M=0$. Since vector chirality is even under the time reversal, finite $M$ is required to induce $\sigma_{xy}$, which is odd under the time reversal. 
\begin{figure}\centering
\includegraphics[scale=1.0]{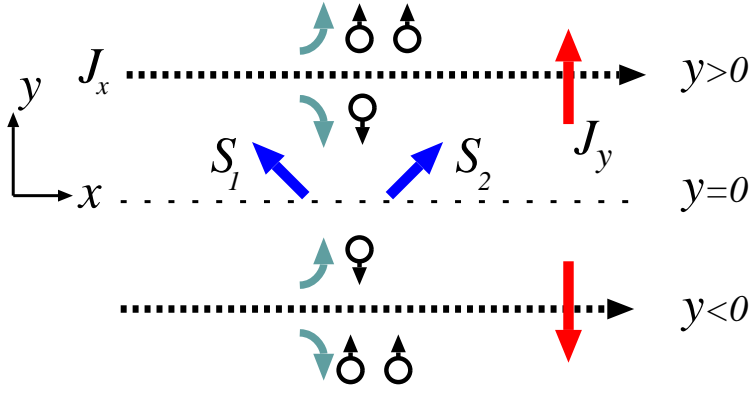}
\caption{The local Hall effect due to the vector chirality, $\bm{S}_{1}\times\bm{S}_{2}$, is due to the local spin Hall effect, which leads to deviate the electron motion in the opposite sense depending on the electron spin (represents by the tiny arrow with circles). 
The spins of the Hall effect are opposite in the regions $y>0$ and $y<0$ due to the different sign of the geometry factor.
Therefore, global Hall effect vanishes if the electron transport is homogeneous. }\label{Fig:2}
\end{figure}
In other words, the Hall effect is due to the local spin Hall effect\cite{rf:Hirsch_99} where the electron orbit is deviated by the vector spin chirality in the opposite sense depending on its spin (Fig. \ref{Fig:2}). 
We also note that uniform Hall conductivity, given as the spatial integral of $\sigma_{xy}(\bm{x})$, vanishes if the electron transport is homogeneous and if the system size infinity. 
In fact, $A_{12}$ has a property of $A_{12}(\bm{x})=-A_{12}(\bm{x'})$ at the position $\bm{x'}=\bm{x}_1+\bm{x}_2-\bm{x}$ and this symmetry cancels out the global Hall conductivity. 
This is explained also in Fig. \ref{Fig:2}. 
When the two local spins are on the $\bm{x}$ axis ($y=0$), the local Hall effects in the regime $y>0$ and $y<0$ have opposite sign due to the opposite sign of the geometrical factor, $(\bm{x}_{10}\times\bm{x}_{20})$ and $(\bm{x}_{12}\times\bm{x}_{10})_{z}$.
Therefore, the global Hall effect vanishes except in the case where the electron transport is asymmetric withe respect to the spin structure. 
 Nevertheless, we believe that finite global Hall conductivity can arise in reality, since electron transport is not necessarily uniform and the dominant transport channel can be affected by a net vector chirality. 

When the localized spins is slowly varying compared to the conduction electrons, i.e., $\ell \gg \lambda$, where $\lambda$ is the spatial scale of the local spin structure, we can expand the local spin as  $\bm{S}_{\bm{x}_1}\simeq\bm{S}_{\bm{x}}+(\bm{x}_{10}\cdot\bm{\nabla}_{\bm{x}})\bm{S}_{\bm{x}}+\cdots$. The Hall conductivity can be then simplified as
\begin{align} \label{eq:AHC of vector chirality} 
	\sigma_{xy}(\bm{x})&=	\alpha(\partial_{\bm{x}}\bm{S}(\bm{x}) \times \partial_{\bm{y}}\bm{S}(\bm{x}) )^z,
\end{align}
with the coefficient $\alpha$ given by
\begin{align}\label{eq:E0}
\alpha &={\frac{i}{2V} \left (\frac{e\hbar}{m} \right)^2 
J_{\mathrm{sd}}^2 }\sum_{k\sigma}\sigma G^R_{k,\sigma}G^R_{k,-\sigma} |G^R_{k,\sigma}|^2
\\ \label{eq:E1}
&\simeq \frac{3\pi}{2}\sigma_{\mathrm{B}}\frac{M J_{\mathrm{sd}}^2\tau}{\hbar}\left(
{\left(\frac{M}{\epsilon_{\mathrm{F}}}\right)^2+\frac{1}{4}\left(\frac{\hbar}{\epsilon_{\mathrm{F}}\tau}\right)^2} \right)^{-1},
\end{align}
where $\sigma_{\mathrm{B}}=\frac{e^2 n_{e} \tau}{m}$ is Boltzmann conductivity and  $n_{e}=\frac{1}{3}\epsilon_{\mathrm{F}} N_{e}$ is the electron density.

The behavior of the coefficient of $\alpha$ depends on the magnitude of the spin polarization, namely,  $\alpha \sim 6\pi\frac{M}{\epsilon_{\mathrm{F}}}\sigma_{\mathrm{B}}$ if $\frac{M\tau}{\hbar}\ll1$, and  $\alpha \sim \frac{3\pi}{2}\frac{\epsilon_{\mathrm{F}}}{M}\sigma_{\mathrm{B}}$ if $\frac{M\tau}{\hbar}\gg1$.   

From Eq. (\ref{eq:AHC of vector chirality}), we see that the anomalous Hall effect is related to the spin current carried by the localized spins. In fact, the local spin current is given in the static case and in the absent of the spin-orbit interaction as\cite{rf:Takeuchi_08},
\begin{align}
J^{z}_{\mathrm{s}, \mu}=\xi \epsilon_{ijz}\bm{S}^i\nabla_{\mu}\bm{S}^j,
\end{align}
where $\xi$ is a constant (the superscript (subscript) of  $J^{z}_{\mathrm{s}, \mu}$ represents the direction in the spin (real) space). 
This spin current is a nonlinear magnetic current of local spins (and is different from the conventional magnetic current in the electromagnetism, $\bm{J}^{\nu}_{\mathrm{M}}=\nabla\times M^{\nu}$). 
The local Hall conductivity of Eq. (\ref{eq:AHC of vector chirality}) is thus written by the rotation of vector spin chirality as $\sigma_{xy}(\bm{x})= {\frac{\alpha}{\xi}} \left( \nabla\times\bm{J}^{z}_\mathrm{s} \right)_{z}$. The global Hall conductivity is then written as 
\begin{align}  \notag
\sigma_{xy} &=\iint \mathrm{d}x \mathrm{d}y  \frac{\alpha}{\xi} \left( \nabla\times\bm{J}^{z}_\mathrm{s} \right)_z
\\ \label{eq:Berry phase curvature} 
&= \frac{\alpha}{\xi}\oint \mathrm{d}\bm{\ell}\cdot \bm{J}^{z}_\mathrm{s}, 
\end{align}
where $\mathrm{d}\bm{\ell}$ denotes the line integral at the boundary. Equation (\ref{eq:Berry phase curvature}) indicates that Hall voltage is induced by the rotation of the spin current.
\begin{figure}\centering
\includegraphics[scale=1.0]{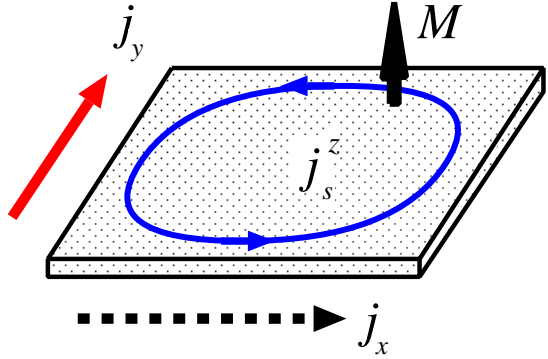}
\caption{The Hall current $j_{y}$ is interpreted as due to the rotational spin current at the boundary associated with the local spin structure.
         }\label{Fig:5}
\end{figure}

\section{Discussion}
The result of the local Hall conductivity, Eq. (\ref{eq:Hall conducutivity-1}), is very similar to the expression of the Dzyaloshinskii-Moriya interaction, 
$\mathcal{H}_{\mathrm{DM}}=\sum_{ij}\bm{D}_{ij}\cdot\left(\bm{S}_{i}\times\bm{S}_{j}\right)$, where $\bm{D}_{ij}$ is a coefficient and $i$, $j$ represent the position.
We therefore expect that the systems having Dzyaloshinskii-Moriya interaction exhibit the Hall effect due to the vector spin chirality as $\sigma_{xy}(\bm{x})\propto A_{12}(\bm{x})D_{12}^z $.
The vector chirality effect would be seen in the system with helical spin structures such as MnSi \cite{rf:Fabris_06} and AuFe \cite{rf:Binz_08}.
\begin{figure}\centering
\includegraphics[scale=1.0]{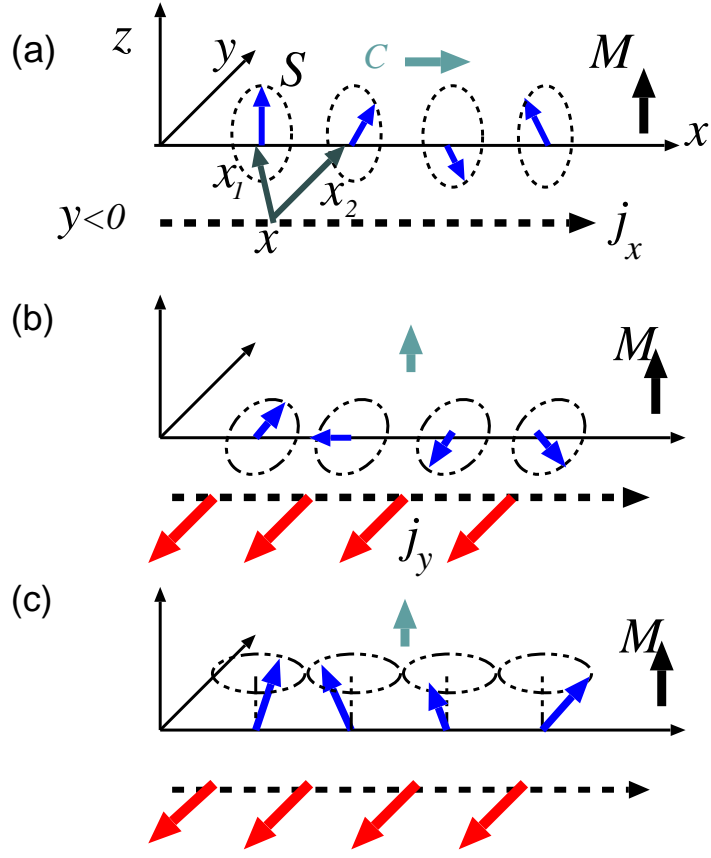}
\caption{The local Hall current in typical spin structures. (a) The helical spin structure in the $y$-$z$ plane. No Hall current arises. (b) The helical spin structure in the $x$-$y$ plane. Local Hall current is induced. (c) The conical spin structure also induces the local Hall current.
         }\label{Fig:3}
\end{figure}

Equation (\ref{eq:Hall conducutivity-1}) has the similar structure as the electric polarization in the magnetic insulator. In fact, the local vector spin chirality was pointed to induce the polarization $\bm{P}$ as $\bm{P}=a(\bm{x}_{2}-\bm{x}_{1})\times(\bm{S}_{1}\times\bm{S}_{2})$, where $a$ is a  constant proportional to the spin-orbit interaction and exchange interaction \cite{rf:Katsura_05,rf:Mostovoy_06,rf:Sergienko_06}. This relation was confirmed experimentally in manganese oxides with helical and conical spin structures\cite{rf:Yamasaki_07}. One should note that there is a difference between the anomalous Hall effect in metals and the spontaneous electric polarization in insulators.
Namely, while the spontaneous polarization arises when the spin-orbit interaction is present, the coefficient for the Hall effect, $A_{12}$ of Eq. (\ref{eq:Hall conducutivity-1}), does not contain the spin-orbit interaction.

Let us apply our result explicitly to the typical spin structures, shown in Fig. \ref{Fig:3}. Figure. \ref{Fig:3}(a) shows the helical spin structure with the spins lying in the plane perpendicular to the current ($x$-direction). The spin is written as $\bm{S}(\bm{x})=S(0, \sin(\bm{q}\cdot\bm{x}), \cos({\bm{q}\cdot\bm{x}}) )$, where $\bm{q}$ represents the pitch of the helical structure. In this case, the vector spin chirality in the $z$-direction, 
$\bm{C}^z=\left(\bm{S}_{1}\times\bm{S}_{2}\right)^z$, vanishes and therefore the Hall effect does not arise. When the helical spins are within the $x$-$y$ plane (Fig. \ref{Fig:3}(b)), spin structure is given as $\bm{S}(\bm{x})=S(\cos(\bm{q}\cdot\bm{x}), \sin({\bm{q}\cdot\bm{x}}), 0)$, and the vector chirality is $\left(\bm{S}_{1}\times\bm{S}_{2}\right)^z=S^2 \sin\left(\bm{q}\cdot\bm{x_{12}}\right)$. Similarly, the conical spin structure shown in Fig. \ref{Fig:3}(c), with $\bm{S}(\bm{x})=S(\cos(\bm{q}\cdot\bm{x}), \sin({\bm{q}\cdot\bm{x}}), S^z)$, ($S^z$ is a constant), results in $\left(\bm{S}_{1}\times\bm{S}_{2}\right)^z=S^2 \sin\left(\bm{q}\cdot\bm{x_{12}}\right)$ and finite local Hall effect arises.
\begin{figure}\centering
\includegraphics[scale=1.0]{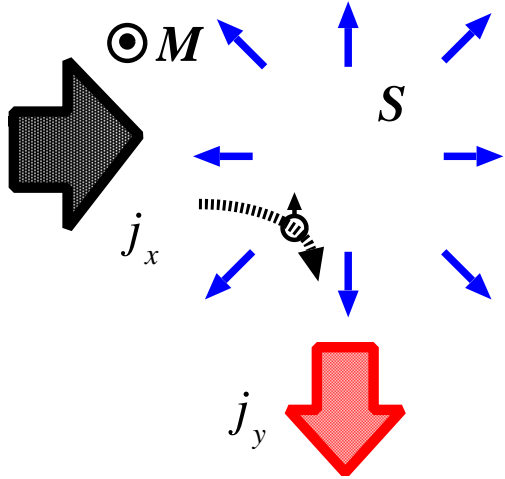}
\caption{The local Hall current caused by a vortex spin structure.
         }\label{Fig:4}
\end{figure}

A typical case where the Hall effect arises in the whole sample is the case with a vortex. For a vortex shown in Fig. \ref{Fig:4}, the spin configuration is given as
	\begin{align}
	\bm{S}={S}(\sin{\theta}\cos{\phi}, \sin{\theta}\sin{\phi}, \cos{\theta}),
	\end{align}
where $\theta, \phi$ represent the azimuthal and  polar angles in polar coordinates. The Hall conductivity in this case is calculated from Eq.(\ref{eq:Berry phase curvature}), 
\begin{align}\label{eq:div} 
\sigma_{xy}=-\pi\frac{\alpha}{\xi}{S^2}N^{\mathrm{vor}}.
\end{align}
The Hall conductivity for the vortex is therefore finite and is proportional to the total vortex number $N^{\mathrm{vor}}$.

Finally, we note that the line integral of the spin current in Eq. (\ref{eq:Berry phase curvature}) is equivalent to the SU(2) gauge flux.  The conventional anomalous Hall conductivity can be induced by the SU(2) gauge flux\cite{rf:Ye_99}. In fact, SU(2) gauge field associated with the spin structure is given as\cite{rf:Tatara_08} $\bm{A}_{\mu}=\frac{1}{2 S^2}\bm{S}\times\nabla_{\mu}\bm{S} - \frac{1}{2S}(1-\cos{\theta})\nabla_{\mu}\phi\cdot\bm{S}$, and therefore $\bm{j}_{s, \mu}\propto \bm{A}_{\mu}^{\perp}$, where $\perp$ indicates the component perpendicular to $\bm{S}$. The Hall conductivity, Eq. (\ref{eq:Berry phase curvature}), is thus written in terms of the perpendicular to component of the SU(2) gauge flux as  
\begin{align}
\sigma_{xy}=\frac{2S^2 \alpha}{\xi}\oint \mathrm{d}\bm{\ell}\cdot (\bm{A}^{\perp})^{z}.
\end{align}
The Hall effect studied here is therefore interpreted as due to an U(1) projection of the SU(2) phase accumulated by the conductive electron. One should note that this projection is different from the scalar spin chirality or the spin Berry phase case. The U(1) phase governing the scalar chirality is given by the parallel component, $\bm{A}^\parallel\equiv(\bm{A}\cdot \bm{S})$, of the gauge field.

\section{Conclusion}
We have shown that the anomalous Hall conductivity is caused by the vector spin chirality at least locally  in the weak $s$-$d$ (Hund) coupling limit if the electron has uniform spin polarization, without the spin-orbit interaction. This anomalous Hall effect can be interpreted as due to rotation of the spin current associated with the vector chirality. The anomalous Hall effect is expected if the edge spin current exists as in the quantum spin Hall systems\cite{rf:Kane_05-1}.

\section*{Acknowledgments}
This work was supported by a Grant-in-Aid for Scientific Research in Priority Area "Creation and control of spin current" (1948027) from the Ministry of Education, Culture, Sports, Science and Technology, Japan.

\end{document}